\documentclass[%
prb
,twocolumn%
,amsmath,amssymb,aps,nobibnotes
,notitlepage
,nofootinbib
,preprintnumbers
]{revtex4-2}

\usepackage[pdftex]{graphicx}
\usepackage[colorlinks=true,linkcolor=blue,urlcolor=blue,citecolor=blue]{hyperref}
\usepackage{xcolor}
\makeatletter
\let\MYcaption\@makecaption
\makeatother

\usepackage{subcaption}
\makeatletter
\let\@makecaption\MYcaption
\makeatother

\usepackage{bm}
\usepackage{braket}
\usepackage{color}
\usepackage{tabularx}
\usepackage{mathtools}
\usepackage{physics}
\usepackage{siunitx}
\newcommand{\imp}{\mathrm{imp}}

\begin{document}
\title{Cavity-enhanced Kondo effect}

\author{Jun Mochida}
\email{jun-mochida@g.ecc.u-tokyo.ac.jp}
\affiliation{Department of Physics, University of Tokyo, 7-3-1 Hongo, Bunkyo-ku, Tokyo 113-0033, Japan}

\author{Yuto Ashida}
\email{ashida@phys.s.u-tokyo.ac.jp}
\affiliation{Department of Physics, University of Tokyo, 7-3-1 Hongo, Bunkyo-ku, Tokyo 113-0033, Japan}
\affiliation{Institute for Physics of Intelligence, University of Tokyo, 7-3-1 Hongo, Tokyo 113-0033, Japan}

\date{\today}

%\pacs{***}

\begin{abstract}
In metals containing magnetic impurities, conduction electrons screen the magnetic impurities and induce the Kondo effect, i.e., the enhancement of the electrical resistance at low temperatures. Motivated by recent advances in manipulating quantum materials by cavity confinement, we study how the ultrastrong light-matter coupling can affect the Kondo effect. 
We show that the ultrastrong coupling can enhance the Kondo temperature and give rise to several notable phenomena, including universal scalings of the cavity-modified Kondo effect, the photon occupation number, and the entanglement entropy between the cavity and electrons. The origin of the cavity enhancement can be understood from the mass renormalization due to the cavity-mediated nonlocal electron-electron interaction, which is akin to the polaronic mass enhancement.  We combine the unitary transformations and the Gaussian variational states to analyze the quantum impurity system confined in the cavity. Our nonperturbative framework can be applied to a variety of quantum impurity problems influenced by structured quantum electromagnetic environment.
\end{abstract}

\maketitle

\section{Introduction\label{sec:intro}}
The Kondo effect is arguably one of the most fundamental themes in condensed matter physics~\cite{kondoResistanceMinimumDilute1964,hewsonKondoProblemHeavy1993,colemanIntroductionManyBodyPhysics2015}. 
When a localized impurity spin is embedded in a Fermi gas, an antiferromagnetic exchange interaction occurs between the impurity and the conduction electrons. 
This interaction causes nonmonotonic temperature dependence in transport, leading to the minimum in electrical resistivity around the Kondo temperature $T_{\rm K}$~\cite{andersonLocalizedMagneticStates1961, kondoResistanceMinimumDilute1964,sarachikResistivityMoNbMoRe1964}. 
At low temperatures $T<T_{\rm K}$, the impurity spin is screened by the surrounding electrons through the antiferromagnetic exchange interaction. The effective interaction strength diverges in the low-energy limit, which gives rise to the singlet ground state called the Kondo singlet state~\cite{sorensenScalingTheoryKondo1996,affleckDetectingKondoScreening2001,affleckKondoScreeningCloud2010,bordaKondoScreeningCloud2007,bordaKondoCloudSpinspin2009}. The size of the screening cloud around the impurity, the Kondo cloud, is characterized by the Kondo length $\xi_\mathrm{K}$, which is typically an order of micrometer~\cite{v.borzenetsObservationKondoScreening2020} and can be related to the Kondo temperature by $\xi_\mathrm{K}=\hbar v_{\rm F}/k_\mathrm{B}T_\mathrm{K}$ with the Fermi velocity $v_{\rm F}$. 

Understanding of the Kondo effect has played a central role in many areas of solid-state physics, such as heavy fermions~\cite{hewsonKondoProblemHeavy1993},  mesoscopic physics \cite{goldhaber-gordonKondoEffectSingleelectron1998,vanderwielKondoEffectUnitary2000,pustilnikKondoEffectQuantum2004,kobayashiShotNoiseMesoscopic2021}, and dynamical mean-field theory~\cite{georgesDynamicalMeanfieldTheory1996}. 
In particular, the internal structures of the Kondo cloud, such as the quantum correlations or entanglement therein, are still under investigations in both theory~\cite{
alkurtassEntanglementStructureTwochannel2016,yangUnveilingInternalEntanglement2017,kimUniversalThermalEntanglement2021,mocaKondoCloudSuperconductor2021}
and experiments~\cite{v.borzenetsObservationKondoScreening2020,imObservationKondoCondensation2023,leeKondoScreeningMajorana2023}. 
Controlling and emulating the Kondo effect have also been widely explored in a variety of setups, including tunable quantum dot systems with an external magnetic field~\cite{goldhaber-gordonKondoEffectSingleelectron1998} or finite-volume confinement called the Kondo box~\cite{thimmKondoBoxMagnetic1999,simonFiniteSizeEffectsConductance2002}, defects in a graphene~\cite{chenTunableKondoEffect2011,fritzPhysicsKondoImpurities2013}, periodically driven materials~\cite{nordlanderKondoPhysicsSingleelectron2000,kaminskiUniversalityKondoEffect2000,heylNonequilibriumSteadyState2010,takasanLaserirradiatedKondoInsulators2017,ecksteinTwochannelKondoPhysics2017,fausewehLaserPulseDriven2020,quitoFloquetEngineeringMultichannel2023}, and ultracold atoms~\cite{duanControllingUltracoldAtoms2004,gorshkovTwoorbitalMagnetismUltracold2010,nakagawaLaserInducedKondoEffect2015,kanasz-nagyExploringAnisotropicKondo2018,nakagawaNonHermitianKondoEffect2018,rieggerLocalizedMagneticMoments2018,onoObservationSpinexchangeDynamics2021}.

On another front, the field of cavity quantum electrodynamics (QED) has long played an important role in quantum technology and quantum information. Cavity QED has traditionally strived to study interaction between electromagnetic fields and matter mainly in few-body regimes. Recently, there have been significant efforts in understanding the possibility of employing the idea of cavity QED to control quantum many-body systems in a stable manner without any external driving~\cite{friskkockumUltrastrongCouplingLight2019,garcia-vidalManipulatingMatterStrong2021,hubenerEngineeringQuantumMaterials2021,schlawinCavityQuantumMaterials2022}. 
Instead of using electromagnetic fields as classical external field, this emerging area of research, namely  cavity QED materials aims to exploit strong couplings between materials and vacuum fluctuations of the quantized electromagnetic field in the cavity.

So far, a number of experiments have realized the so-called ultrastrong coupling regime in a variety of setups~\cite{anapparaSignaturesUltrastrongLightmatter2009,forn-diazUltrastrongCouplingRegimes2019,kellerLandauPolaritonsHighly2020,kippCavityElectrodynamicsVan2024}, where the light-matter coupling strength is comparable to elementary excitation energies. 
Previous studies have explored the  possibility of employing such ultrastrong light-matter coupling to control excitations~\cite{smolkaCavityQuantumElectrodynamics2014,scalariSuperconductingComplementaryMetasurfaces2014,zhangCollectiveNonperturbativeCoupling2016,bayerTerahertzLightMatter2017,hagenmullerCavityEnhancedTransportCharge2017,bartoloVacuumdressedCavityMagnetotransport2018,rokajQuantumElectrodynamicalBloch2019,liElectromagneticCouplingTightbinding2020,halbhuberNonadiabaticStrippingCavity2020,ashidaCavityQuantumElectrodynamics2023,herzigsheinfuxHighqualityNanocavitiesMultimodal2024,roman-rocheEffectiveTheoryMatter2022,masukiCavityMoirMaterials2023} and certain material properties,  such as  superconductivity~\cite{sentefCavityQuantumelectrodynamicalPolaritonically2018,schlawinCavityMediatedElectronPhotonSuperconductivity2019,curtisCavityQuantumEliashberg2019a,gaoHiggsModeStabilization2021}, ferroelectricity~\cite{ashidaQuantumElectrodynamicControl2020,latiniFerroelectricPhotoGround2021,lenkCollectiveTheoryInteracting2022,lenkDynamicalMeanfieldStudy2022}, band topology~\cite{appuglieseBreakdownTopologicalProtection2022,wangCavityQuantumElectrodynamical2019a,ciutiCavitymediatedElectronHopping2021,rokajPolaritonicHofstadterButterfly2022,masukiBerryPhaseTopology2023,shafferEntanglementTopologySuSchriefferHeeger2023}, transport \cite{appuglieseBreakdownTopologicalProtection2022,jarcCavitymediatedThermalControl2023}, and chemical reactivity~\cite{herreraCavityControlledChemistryMolecular2016,flickAtomsMoleculesCavities2017,ruggenthalerQuantumelectrodynamicalLightMatter2018,ribeiroPolaritonChemistryControlling2018}. 
There, excited states and even ground-state electronic properties can be modified due to virtual processes in which both matter and cavity photons are excited. At terahertz frequencies, which are typically relevant to excitations in real materials, the ultrastrong coupling regime has been so far achieved with the collective enhancement, where the light-matter coupling is enhanced by a factor $\sqrt{N}$ with $N$ being the number of elements coupled to the cavity mode. 
In such regime, common simplifications in cavity QED fail, rendering theoretical analysis challenging. For instance, rotating wave approximations and/or an effective description based on projection onto low-energy manifold, such as the two-level approximation and a tight-binding description, can in general no longer be justified. Thus, to accurately analyze cavity QED materials in the ultrastrong coupling regime,  one needs to employ a nonperturbative method that does not rely on  uncontrolled simplifications.

\begin{figure}[t]
  \centering
    \includegraphics[width=1\columnwidth]{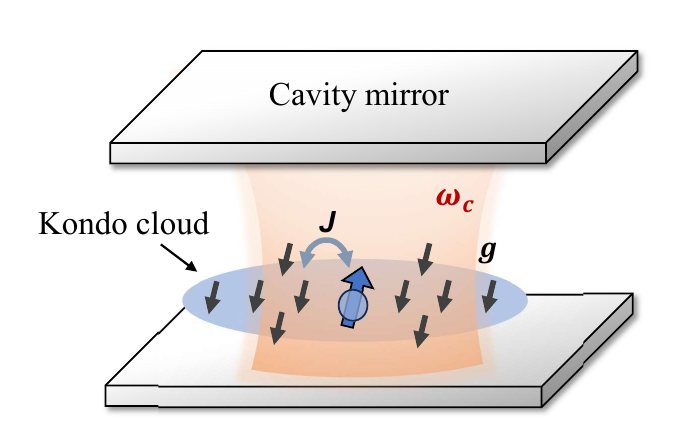}
    \caption{Schematic figure of the Kondo singlet confined in a cavity. A metal including the magnetic impurity is embedded in the cavity, and the quantized electromagnetic field at frequency $\omega_c$ couples to the conduction electrons, where the coupling strength is denoted by $g$. The magnetic impurity is screened by the conduction electrons through the exchange interaction $J$ and forms the Kondo screening cloud.}
	\label{fig:cavityKondo}
\end{figure}

Motivated by these developments, in this paper we study how the Kondo effect can be influenced by the ultrastrong coupling to the quantum electromagnetic field confined in the cavity (Fig.~\ref{fig:cavityKondo}). To analyze the problem in an efficient and accurate way, we employ the two unitary transformations to firstly asymptotically decouple the electronic system from cavity photons ~\cite{ashidaCavityQuantumElectrodynamics2021,ashidaNonperturbativeWaveguideQuantum2022} and secondly completely disentangle the localized impurity and the conduction electrons \cite{ashidaSolvingQuantumImpurity2018,ashidaVariationalPrincipleQuantum2018,kanasz-nagyExploringAnisotropicKondo2018}. We then obtain an effective model that can capture low-energy physics of the cavity Kondo effect, where the leading contribution due to the cavity confinement emerges as the nonlocal electron-electron interaction mediated by cavity photons. We analyze the ground-state properties of this effective model by using fermionic Gaussian variational states. The results indicate that the nonlocal interaction effectively increases the density of states near the Fermi sea, thus enhancing the Kondo temperature $T_{\rm K}$, which is akin to the polaronic mass enhancement. We also find that the ultrastrong coupling leads to universal scalings of the cavity-modified Kondo temperature, the photon occupation number, and the entanglement entropy between the cavity and electrons as a function of the light-matter coupling strength $g$ scaled by $T_{\rm K}$.  

We note that our study makes a contrast to existing studies on the related topics. The effect of cavity confinement has been discussed in the setup of a quantum dot connected with external reservoirs~\cite{kuoKondoQEDKondo2023b}. There, the cavity field does not couple to conduction electrons, but only couples to {\it impurity sites} in the quantum dot. It has been argued that the cavity field, which directly perturbs the localized impurity, inhibits the formation of the Kondo state and suppresses the Kondo effect in this case. In contrast, our work focuses on the setup relevant to recent experiments of cavity QED materials, namely,  solid-state material embedded in, e.g, the plasmonic cavity, where the cavity field couples to {\it conduction electrons} in bulk. In the present case, the cavity confinement is shown to induce the opposite behavior, i.e., it enhances the Kondo effect. Meanwhile, Ref.~\cite{mullerControlYuShibaRusinovStates2023} has discussed the possibility of controlling the Yu-Shiba-Rusinov state by the coupling to a bosonic mode on the basis of the Peierls substitution. The results suggest that the coupling to a bosonic mode can modify the strength of the exchange interaction of the magnetic impurity embedded in a conventional superconductor. Our analysis is complementary to this previous work in the sense that we consider the magnetic impurity embedded in a normal metal, and we treat the light-matter coupling nonperturbatively without resorting to uncontrolled approximations such as the Peierls substitution, which can break down in the ultrastrong coupling regime \cite{masukiBerryPhaseTopology2023}.

The rest of the paper is organized as follows. 
In Sec.~\ref{sec:model_and_method}, we introduce a model for the magnetic impurity in a metal confined in the cavity and use the unitary transformations to derive an effective single-impurity model to describe the low-energy physics. 
We also explain the non-Gaussian variational method that can be used to study the ground-state properties of the model.  In Sec.~\ref{sec:result}, we present the variational results that indicate the cavity-enhanced Kondo effect. 
We also discuss the emergence of new types of universal relations in the cavity-enhanced Kondo effect. 
Finally, in Sec.~\ref{sec:discuss}, we discuss the understanding of the cavity-enhanced Kondo effect using poor person's scaling and summarize the results. This section also discusses future perspectives and possible experimental relevance.

\section{Model and method\label{sec:model_and_method}}

\subsection{Magnetic impurity embedded in cavity\label{sec:model}}
We derive an effective low-energy model of the Kondo effect in the presence of the ultrastrong light-matter coupling. To this end, we start from the single-impurity Anderson model with light-matter interaction. Specifically, we assume the long-wave approximation, i.e., neglect the spatial dependence of the cavity field,  and consider  the following light-matter Hamiltonian in the Coulomb gauge
\begin{align}
	\hat{H}&=\hat{H}_{\mathrm{e}}+\hat{H}_{\imp}+\hat{H}_{\mathrm{cavity}}, 
\end{align}
where  $\hat{H}_\mathrm{cavity}=\hbar\omega_\mathrm{c}\hat{a}^{\dagger}\hat{a}$ describes the energy of the single-mode cavity photon with the cavity frequency $\omega_c$, and the photon annihilation (creation) operator $\hat{a}$ ($\hat{a}^\dagger$) satisfies the canonical commutation relation $[\hat{a},\hat{a}^\dagger]=1$. 
The first term $\hat{H}_\mathrm{e}$ represents a one-dimensional electron system with the light-matter coupling,
\begin{align}\label{eq:anderson-cqed}
	\hat{H}_{\mathrm{e}}&=%\sum_{k,\sigma}\frac{\bigl(\hbar k-\bm{A}\bigr)^2}{2m}\hat{c}^\dagger_{k\sigma}\hat{c}_{k\sigma}
	\sum_{\sigma}\int dx\,\hat{\psi}_\sigma^\dagger(x)\frac{\bigl(\hat{p}+e\hat{A}\bigr)^2}{2m}\hat{\psi}_\sigma(x)
\end{align}
with $\hat{A}=A(\hat{a}+\hat{a}^\dagger)$ being the vector potential and
\begin{align}
	\hat{\psi}_\sigma(x)=\frac{1}{\sqrt{L}}\sum_{k}e^{ikx}\hat{c}_{k\sigma},
\end{align} 
where $L$ is the system size and $\hat{c}_{k\sigma}\ (\hat{c}^\dagger_{k\sigma})$ is the electron annihilation (creation) operator with momentum $k$ and spin $\sigma\in\{\uparrow,\downarrow\}$, while $\hat{p}=-i\hbar\partial_x$ is the momentum operator. We impose the periodic boundary conditions such that $k_n=2\pi n/L,\ n=0,\pm1,\pm2,\dots,\pm N_\Lambda$ with a cutoff integer number $N_\Lambda$. The constant $m$ is an effective electron mass, $e$ is the elementary charge, and $N_\mathrm{e}$ is the total number of conduction electrons. 
The impurity Hamiltonian is 
\begin{align}
	 \hat{H}_{\imp}&=\sum_{\sigma}\varepsilon_{d}\,\hat{d}^\dagger_{\sigma}\hat{d}_\sigma+U\hat{n}_{d\uparrow}\hat{n}_{d\downarrow}\notag 
	\\ &\hspace{20mm}+\frac{V_0}{\sqrt{N_\mathrm{e}}}\sum_{k,\sigma} \qty(\hat{d}^{\dagger}_{\sigma}\hat{c}_{k\sigma}+\hat{c}_{k\sigma}^\dagger \hat{d}_{\sigma}),
	\label{eq:imp}
\end{align}
where $\varepsilon_d$ is the impurity energy, $\hat{d}_\sigma$ ($\hat{d}_\sigma^\dagger$) annihilates (creates) an electron at the localized impurity site with spin $\sigma$, and $U$ denotes the Coulomb repulsion at the impurity site with the number operator $\hat{n}_{d\sigma}=\hat{d}_\sigma^\dagger\hat{d}_\sigma$. We note that the fermion operators $\hat{c}_{k\sigma}$ and $\hat{d}_\sigma$ satisfy the canonical anti-commutation relations
\begin{gather}
	\{\hat{c}_{k\sigma},\hat{c}_{k'\sigma'}^\dagger\}=\delta_{kk'}\delta_{\sigma\sigma'},\
	\{\hat{d}_{\sigma},\hat{d}_{\sigma'}^\dagger\}=\delta_{\sigma\sigma'}.
\end{gather}
The last term in Eq.~\eqref{eq:imp} is the hybridization term between the impurity and the conduction electrons with the hybridization strength $V_0$.

To obtain a low-energy effective model, we firstly use the Bogoliubov transformation to diagonalize the photon-only part of the total Hamiltonian, which includes $\hat{H}_{\rm cavity}$ and the $\hat{A}^2$ term in $\hat{H}_\mathrm{e}$, thereby introducing another boson operator $\hat{b}$ as
\begin{align}\label{eq:bogoliubov}
	\mqty(\hat{b}\\ \hat{b}^\dagger)&=\mqty(\cosh{r} & \sinh{r} \\ \sinh{r}&\cosh{r})\mqty(\hat{a}\\ \hat{a}^\dagger), \\
	r&=\log\qty[\sqrt{\frac{\alpha+1}{2}}+\sqrt{\frac{\alpha-1}{2}}],\\
	\alpha&=\frac{\omega_{\mathrm{c}}}{\Omega}\qty(1+N_\mathrm{e}\frac{e^2A^2}{m\hbar\omega_\mathrm{c}}),
\end{align}
where
\begin{align}
	\Omega=\omega_{\mathrm{c}}\sqrt{1+2N_\mathrm{e}\frac{g^2}{\omega_c^2}}
\end{align}
is the renormalized photon frequency, $g=eA\sqrt{\omega_{\rm c}/(m\hbar)}$ is the strength of the light-matter coupling.

\if0
After this transformation, the Hamiltonian of light-matter systems rewrite as
\begin{align}
    &\hat{H}_\mathrm{e}+\hat{H}_\mathrm{light}\notag\\
    &=\sum_{k,\sigma}\frac{\hbar^2 k^2}{2m}\hat{c}^\dagger_{k\sigma}\hat{c}_{k\sigma}-\frac{eA}{m}\sqrt{\frac{\omega_\mathrm{c}}{\Omega}}\hat{P}_\mathrm{e}(\hat{b}^\dagger+\hat{b})
    +\hbar\Omega \hat{b}^\dagger\hat{b},
    \label{eq:e-light}
\end{align}
\fi

In the Coulomb gauge, an analysis of cavity QED materials becomes challenging in the ultrastrong coupling regime due to the strong electron-photon entanglement; the latter leads to the need of including high-energy levels of elementary excitations, such as the high electron bands and bosonic Fock states with large photon occupation numbers \cite{debernardisBreakdownGaugeInvariance2018,liElectromagneticCouplingTightbinding2020,ashidaCavityQuantumElectrodynamics2021}. To overcome such difficulty, we utilize the asymptotically decoupling (AD) unitary transformation~\cite{ashidaCavityQuantumElectrodynamics2021,ashidaNonperturbativeWaveguideQuantum2022},
\begin{align}\label{eq:AD}
	\hat{U}_\mathrm{AD}&=e^{-i\xi_g\frac{\hat{P}_\mathrm{e}}{\hbar}\cdot i(\hat{b}^\dagger-\hat{b})}, \\
	\xi_g&=\sqrt{\frac{\hbar}{m\omega_c}}\frac{g}{\omega_c(1+2N_\mathrm{e}\frac{g^2}{\omega_c^2})^{\frac{3}{4}}},
\end{align}
where $\hat{P}_\mathrm{e}=\sum_{k,\sigma}\hbar k\hat{c}^\dagger_{k\sigma}\hat{c}_{k\sigma}$ represents the total momentum operator of the conduction electrons. After the transformation, the length scale $\xi_g$ characterizes the effective coupling strength in the new reference of frame as shown below. 
This transformation can mitigate the entanglement between electrons and the cavity field and, in particular, completely disentangle them in the strong-coupling limit because $\xi_g$ vanishes when $g\to\infty$. Since $\xi_g\propto g$ at weak $g$, the coefficient $\xi_g$ remains small over the whole range of $g$, which allows us to perform the perturbative analysis with respect to the term $\xi_{g}k$. 

The resulting Hamiltonian after the unitary transformation is
\begin{align}
    &\hat{H}_\textrm{AD}\notag\\
	&=\hat{U}^\dagger_\mathrm{AD}\hat{H}\hat{U}_\mathrm{AD} \notag\\
    &=\sum_{k,\sigma}\varepsilon_k\hat{c}^\dagger_{k\sigma}\hat{c}_{k\sigma}+\hbar\Omega\hat{b}^\dagger\hat{b}
    -\frac{\hbar^2g^2}{m\Omega^2}\hat{P}_\mathrm{e}^2
    +\sum_{\sigma}\varepsilon_{d}\,\hat{d}^{\dagger}_{\sigma}\hat{d}_{\sigma}\notag\\
    &+U\hat{n}_{d\uparrow}\hat{n}_{d\downarrow}+\frac{V_0}{\sqrt{N_\mathrm{e}}}\sum_{k,\sigma} \qty(e^{\xi_gk(\hat{b}^\dagger-\hat{b})}\hat{d}^{\dagger}_{\sigma}\hat{c}_{k\sigma}+\mathrm{h.c.}),
    \label{eq:anderson-cqed2}
\end{align}
where $\varepsilon_k=\frac{\hbar^2 k^2}{2m}$ is the electron dispersion.
In the limit $U\gg V_0$, we arrive at an effective model of the cavity Kondo effect (see Appendix~\ref{app:exchange} for the derivation),
\begin{align}
	\hat{H}_\mathrm{cK}&=\sum_{k,\sigma}\varepsilon_k\hat{c}^\dagger_{k\sigma}\hat{c}_{k\sigma}
	+J\vec{s}(0)\cdot\vec{S}_\mathrm{imp}-\frac{\hbar^2g^2}{m\Omega^2}\hat{P}_\mathrm{e}^2,
	\label{eq:ckondo}
\end{align}
where $J=4V_0^2/U$ is the Kondo exchange interaction. Here, $\vec{S}_\imp=\vec{\sigma}_\imp/2$ is the spin-$\frac{1}{2}$ operator of the magnetic impurity and $\vec{s}(0)=\sum_{k}\hat{c}^\dagger_{k\sigma}(\vec{\sigma})^{\sigma\sigma'}\hat{c}_{k'\sigma'}/2$ is the electron spin at $x=0$ with a vector of the Pauli matrices $\vec{\sigma}=(\sigma^x,\sigma^y,\sigma^z)$. 
We emphasize that this Hamiltonian no longer contains cavity photons, and the leading contribution from the light-matter interaction appears as the nonlocal electron-electron interaction proportional to $\hat{P}_{\rm e}^2$, which is mediated by the cavity field. Said differently, in this transformed frame, the ground state can be given by a product state of the photon vacuum and the many-electron ground state of $\hat{H}_\mathrm{cK}$.
The cavity-mediated nonlocal interaction effectively increases the mass of the electrons, where the electrons are dressed by a cloud of virtual photons~\cite{ashidaCavityQuantumElectrodynamics2021,masukiBerryPhaseTopology2023} in a manner akin to the polaronic mass enhancement \cite{leeMotionSlowElectrons1953}.  As shown below, such mass renormalization will enhance the density of states $\rho_F$ on the Fermi surface, which scales as $\rho_F\propto m$, leading to the higher Kondo temperature $T_\mathrm{K}\propto e^{-\frac{1}{J\rho_F}}$ and thus the cavity-enhanced Kondo effect. 

We note that the ground-state expectation value of the nonlocal interaction, which is proportional to $\expval*{\hat{P}^2_\mathrm{e}}$, should vanish if the total system has the translational symmetry. 
Thus, in the absence of the impurity $J=0$, the ground state is simply a Fermi sea with ${\hat{P}_\mathrm{e}}=0$.
The AD transformation~\eqref{eq:AD} acting on this state reduces to the identity operator, and 
 the light-matter interaction only appears as the squeezing effect induced by the $\hat{A}^2$ term, which is captured by the Bogoliubov transformation~\eqref{eq:bogoliubov}.
In contrast, the presence of the localized impurity breaks the translational symmetry, which renders the effect of the nonlocal interaction $\hat{P}^2_\mathrm{e}$ nontrivial even in the ground state.

\subsection{Variational analysis of the cavity Kondo effect\label{sec:method}}
To analyze the ground-state properties of the Hamiltonian~\eqref{eq:ckondo}, 
we employ the non-Gaussian variational method combining the unitary transformation and a many-body fermionic Gaussian state. 
The fermionic Gaussian wavefunction defines a family of efficient variational  states, where the number of variational parameters increase polynomially as $\order{L^2}$~\cite{krausGeneralizedHartreeFock2010,mitroyTheoryApplicationExplicitly2013}, while it alone cannot capture the strong correlation between the localized impurity and conduction electrons as realized in the Kondo state. 
To overcome this limitation, one can use a unitary transformation to make a larger family of variational states, allowing for efficient and flexible variational calculations.

Specifically, we use the following unitary transformation to completely disentangle the localized impurity and  electrons~\cite{ashidaSolvingQuantumImpurity2018,ashidaVariationalPrincipleQuantum2018}: 
\begin{align}
	\hat{U}_\mathrm{K}=e^{\frac{i\pi}{4}\hat{\sigma}^y_\imp\hat{P}_\mathrm{bath}},
\end{align}
where $\hat{P}_\mathrm{bath}=\exp[{i\pi\sum_{k}\hat{c}^\dagger_{k\uparrow}\hat{c}_{k\uparrow}}]$ is the parity operator of the conduction electrons and $\hat{\sigma}_\imp^\alpha$ with $\alpha=x,y,z$ is the Pauli matrix of the impurity spin.
After the unitary transformation, the impurity can be decoupled from the electrons since the transformed Hamiltonian commutes with the impurity spin, $[\hat{U}^\dagger_\mathrm{K}\hat{H}_\mathrm{cK}\hat{U}_\mathrm{K},\hat{\sigma}^x_\imp]=0$. This fact can be inferred from the parity symmetry of the original Hamiltonian, $[\hat{H}_\mathrm{cK},\hat{\mathbb{P}}]=0$ with $\hat{\mathbb{P}}=\hat{\sigma}^z_\mathrm{imp}\hat{P}_\mathrm{bath}$ satisfying the relation $\hat{U}_\mathrm{K}^\dagger\hat{\mathbb{P}}\hat{U}_\mathrm{K}=\hat{\sigma}^x_\mathrm{imp}$.

Consequently, the impurity spin is no longer a dynamical degree of freedom in the transformed Hamiltonian. 
At the cost of decoupling the impurity, the Hamiltonian acquires the additional nonlocal electron-electron interaction, which originates from the impurity-mediated interaction.  
One can efficiently analyze this transformed Hamiltonian by using variational Gaussian states to study the ground-state properties, which can provide accurate results comparable to tensor network calculations with much less variational parameters \cite{ashidaVariationalPrincipleQuantum2018}.  
We note that it has been also demonstrated that the cavity-mediated nonlocal term $\hat{P}_\mathrm{e}^2$ can be well described by the Gaussian variational states in the context of polaron problems~\cite{dolgirevEmergenceSharpQuantum2021,quEfficientVariationalApproach2022}.

More specifically, as a variational state for the transformed Hamiltonian $\hat{U}^\dagger_\mathrm{K}\hat{H}_\mathrm{cK}\hat{U}_\mathrm{K}$, we use a fermionic Gaussian state $\ket{\psi_\mathrm{GS}}$,
\begin{align}
	\ket{\psi_\mathrm{GS}}&:=e^{\frac{1}{4}\hat{\bm{\psi}}^T X\hat{\bm{\psi}}}\ket{0},
\end{align}
where $X$ is a $2N_f\times 2N_f$ real antisymmetric matrix with $N_f$ being the total number of fermionic modes, and $\ket{0}$ is the vacuum state of conduction electrons; in the present case, $N_f=g_s(2N_\Lambda+1)$ where $g_s=2$ counts the spin degrees of freedom.  
The vector $\hat{\bm{\psi}}$ in the Majorana basis is defined by
\[\hat{\bm{\psi}}=(\hat{\psi}_{1,1},\dots,\hat{\psi}_{1,N_f},\hat{\psi}_{2,1},\dots,\hat{\psi}_{2,N_f})^T\]
with
\begin{align}
	\hat{\psi}_{1,i}&=\hat{c}^\dagger_i+\hat{c}_i,\quad \hat{\psi}_{2,i}=i(\hat{c}^\dagger_i-\hat{c}_i)
\end{align}
for $i=1,2,\dots,N_f$. The fermionic Gaussian wavefunction is completely characterized by its covariant matrix
\begin{align}
	(\Gamma_\psi)_{\eta,\xi}=\frac{i}{2}\expval{[\hat{\psi}_\eta,\hat{\psi}_\xi]}_\mathrm{GS},
\end{align}
where $\expval{\cdot}_\mathrm{GS}$ means the expectation value with respect to a Gaussian state $\ket{\psi_\mathrm{GS}}$.

To obtain the variational ground state, we employ the imaginary-time evolution that minimizes the variational energy~\cite{shiVariationalStudyFermionic2018}
\begin{align}\label{eq:imaginary-time2}
	\dv{\Gamma_\psi}{\tau}&=-\mathcal{H} -\Gamma_{\psi}\mathcal{H}\Gamma_{\psi}, \\ 
	\mathcal{H}:=&4\fdv{E_{\mathrm{var}}}{\Gamma_\psi},\quad E_{\mathrm{var}}=\expval*{{\hat{U}^\dagger_\mathrm{K}\hat{H}_\mathrm{cK}\hat{U}_\mathrm{K}}}_\mathrm{GS}.
\end{align}
This imaginary-time evolution allows us to obtain an approximate ground state within the subspace of the Hilbert space spanned by the  variational wavefunction in the limit $\tau\rightarrow\infty$.
We note that during the imaginary-time evolution the total number of the conduction electrons $N_\mathrm{e}$ is conserved. 
The higher-order many-body correlation function for Gaussian states can be decomposed into components of a covariance matrix by Wick's theorem.
Thus, an expectation value with respect to the Gaussian variational state can be efficiently  calculated for various physical quantities such as magnetic susceptibility and Kondo length as discussed below.
 
It is worthwhile to recall that, in the original Coulomb gauge, the above procedure is  equivalent to finding the variational ground state within the manifold spanned by
\begin{align}\label{gscoulomb}
	\ket*{\mathrm{\psi}}=\hat{U}_\mathrm{AD}\hat{U}_\mathrm{K}\qty(\ket{0}_\mathrm{ph}\ket{\sigma_x}_\imp \ket{\psi_\mathrm{GS}}),
\end{align}
where $\ket{0}_\mathrm{ph}$ is the vacuum state for the photon operator $\hat{b}$, which is a squeezed state in terms of the original photon operator $\hat{a}$, and $\ket{\sigma_x}_\imp$ is the eigenstate of the impurity $x-$spin operator $\hat{\sigma}^x_\imp$ that has eigenvalues $\sigma_x=\pm$. It is evident that the photon-electron entanglement (impurity-electron entanglement) is solely generated by the unitary transformation $\hat{U}_{\rm AD}$ ($\hat{U}_{\rm K}$).

\section{Results}\label{sec:result}
\subsection{Cavity-enhanced Kondo effect}
\begin{figure*}[t]
	\centering
	\includegraphics[width=150mm]{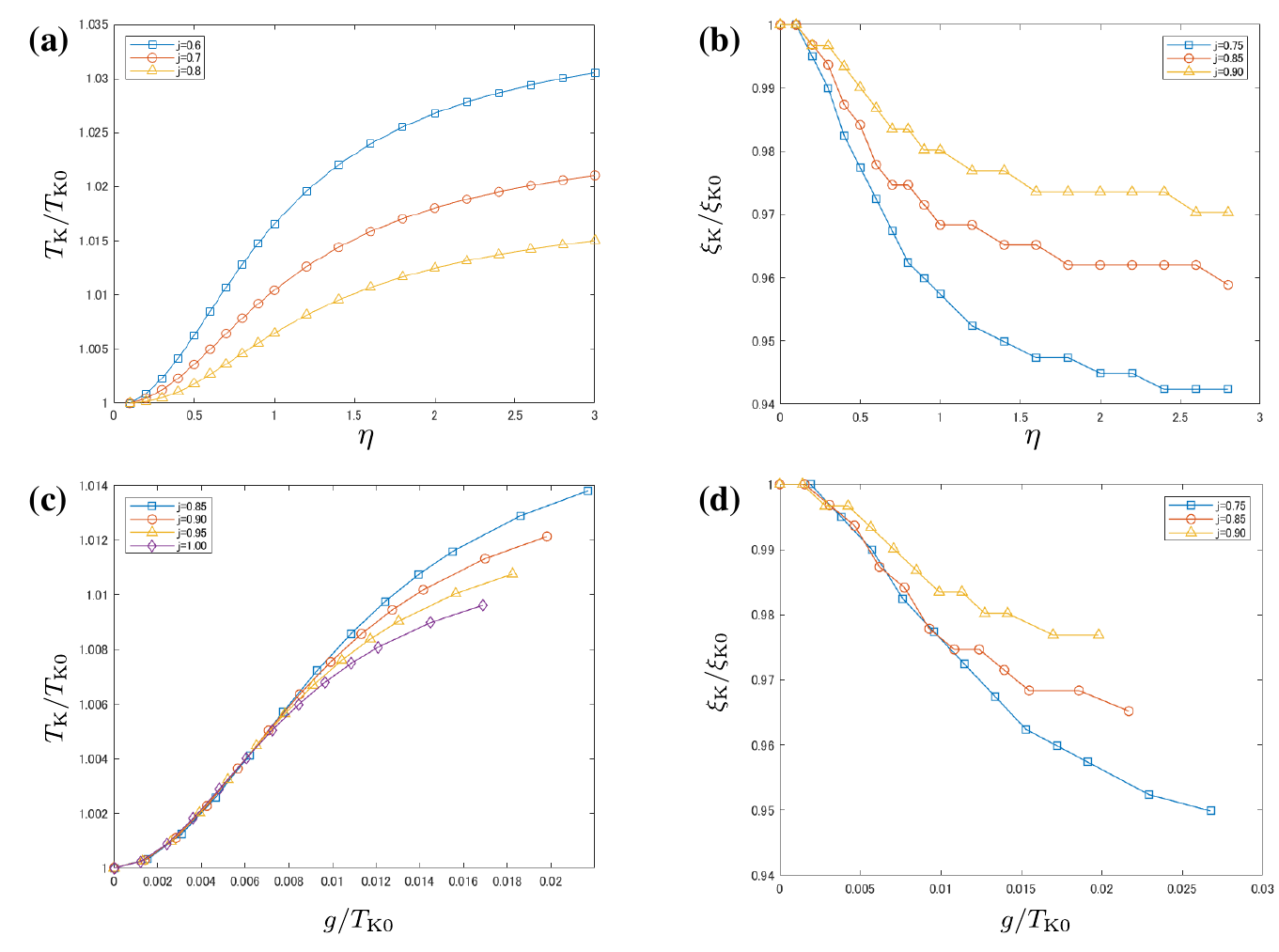}
	\caption{(a) Kondo temperature $T_\mathrm{K}=1/(4\chi_h)$ determined from the magnetic susceptibility of the impurity $\chi_h$ plotted against the dimensionless light-matter coupling $\eta=\sqrt{N_\mathrm{e}}g/\omega_c$.  The Kondo temperature at $\eta=0$ is represented by $T_{\mathrm{K}0}$.
	(b) Kondo length $\xi_\mathrm{K}$ obtained by estimating the distance from the impurity at which the spin correlation is screened as $\int^{\xi_\mathrm{K}}_0 dx \expval*{\vec{s}_\imp\cdot\vec{S}(x)}= -\frac{3}{4}(1-f)$ with $1-f=0.97$. The Kondo length at $g=0$ is represented by $\xi_{\mathrm{K}0}$.
	(c) Kondo temperature plotted against the light-matter coupling strength $g$ normalized by  $T_{\mathrm{K}0}$. The plot covers the range of $\eta\leq1$.
	(d) Kondo length plotted against the light-matter coupling strength $g$ normalized by  $\xi_{\mathrm{K}0}$. 
	The plot covers the range of $\eta\leq0.7$.
	The calculations are performed for $N_\mathrm{e}=70$ in (a,c) and $N_\mathrm{e}=122$ in (b,d). 
	}
	\label{fig:2}
\end{figure*}

We numerically demonstrate that the Kondo effect can be enhanced by the cavity confinement. 
In this section, we use the dimensionless exchange interaction $j=\rho_F J$ with the density of states at the Fermi energy $\rho_F$, and represent the dimensionless light-matter coupling strength via $\eta=\sqrt{N_\mathrm{e}}g/\omega_c$ by including the collective factor $\sqrt{N_{\rm e}}$.
In Fig.~\ref{fig:2}, we show the numerical results of the Kondo temperature extracted from the magnetic susceptibility $\chi_h$ via $T_{\rm K}=1/(4\chi_{h})$ and the Kondo length $\xi_\mathrm{K}$ estimated by the impurity-electron spin correlation function. 
More specifically, the Kondo length is defined as a length scale such that the singlet sum rule~\cite{bordaKondoScreeningCloud2007,holznerKondoScreeningCloud2009}
\begin{align}
	\int dx \expval*{\vec{S}_\imp\cdot\vec{s}(x)}= -\frac{3}{4}
\end{align}
is almost satisfied (see the caption). 
This relation follows from the fact that the total spin operator $\vec{S}_\mathrm{tot}=\vec{S}_\imp+\int dx \vec{s}(x)$ of the Kondo singlet state satisfies the relation $\vec{S}_\mathrm{tot}^2=0$.
All the results are plotted by setting $\hbar=k_\mathrm{B}=m=\omega_c=1$. 

Figure~\ref{fig:2} (a) and (b) show the results of $\eta$ dependence of the Kondo temperature and  length, respectively, where $T_{{\rm K}0}$ and $\xi_{{\rm K}0}$ are the Kondo temperature and length at $\eta=0$.
These numerical calculations consistently indicate that the Kondo effect is enhanced by the cavity confinement, where the Kondo cloud shrinks due to the effectively enhanced exchange interaction $j$.  Physically, the enhancement originates from the cavity-mediated nonlocal interaction proportional to $\hat{P}_\mathrm{e}^2$, which leads to the mass renormalization that enhances the density of states. In the original frame, this phenomenon can be understood as the dressing of conduction electrons by virtual photons, leading to the effect akin to the polaronic mass enhancement.  We note that the Kondo temperature is more sensitive to the change of the dimensionless light-matter coupling strength $\eta$ at a smaller exchange interaction $j$. 
This finding might be understood as follows: a smaller exchange interaction leads to the formation of a more spatially extended Kondo cloud, and the larger number of localized electrons can be susceptible to the cavity confinement effect.

One of the important features of the Kondo effect is that various quantities exhibit universal scaling with the Kondo temperature. 
It is natural to ask whether a similar scaling relation can be found in the cavity Kondo effect discussed here. To examine this possibility, we plot the Kondo temperature and length as a function of the light-matter coupling strength $g$ normalized by $T_{{\rm K}0}$ as shown in Fig.~\ref{fig:2} (c) and (d). 
These numerical results lie on the same universal curves independent of the exchange interaction $j$  up to $\eta\sim1$, suggesting that the cavity-enhancement of the Kondo temperature can exhibit the universal scaling. Thus, the results indicate the scaling relation of the cavity-enhanced Kondo temperature,
\begin{align}\label{eq:temperature_uni}
	\frac{T_\mathrm{K}(j,g)}{T_\mathrm{K0}(j)}=1+f\left(\frac{g}{T_\mathrm{K0}(j)}\right)
\end{align}
with a scaling function $f(x)$ satisfying $f(0)=0$; we recall that $T_{\mathrm{K}0}(j)=T_{\mathrm{K}}(j,0)$ is the Kondo temperature at $g=0$. 
The relation suggests that, even though the light-matter coupling strength $g$ presents as an additional parameter, this parameter can be renormalized in the universal relation of the Kondo effect.

\subsection{Virtual photons induced by the Kondo effect}
\begin{figure}[tb]
	\centering
		\includegraphics[width=8cm]{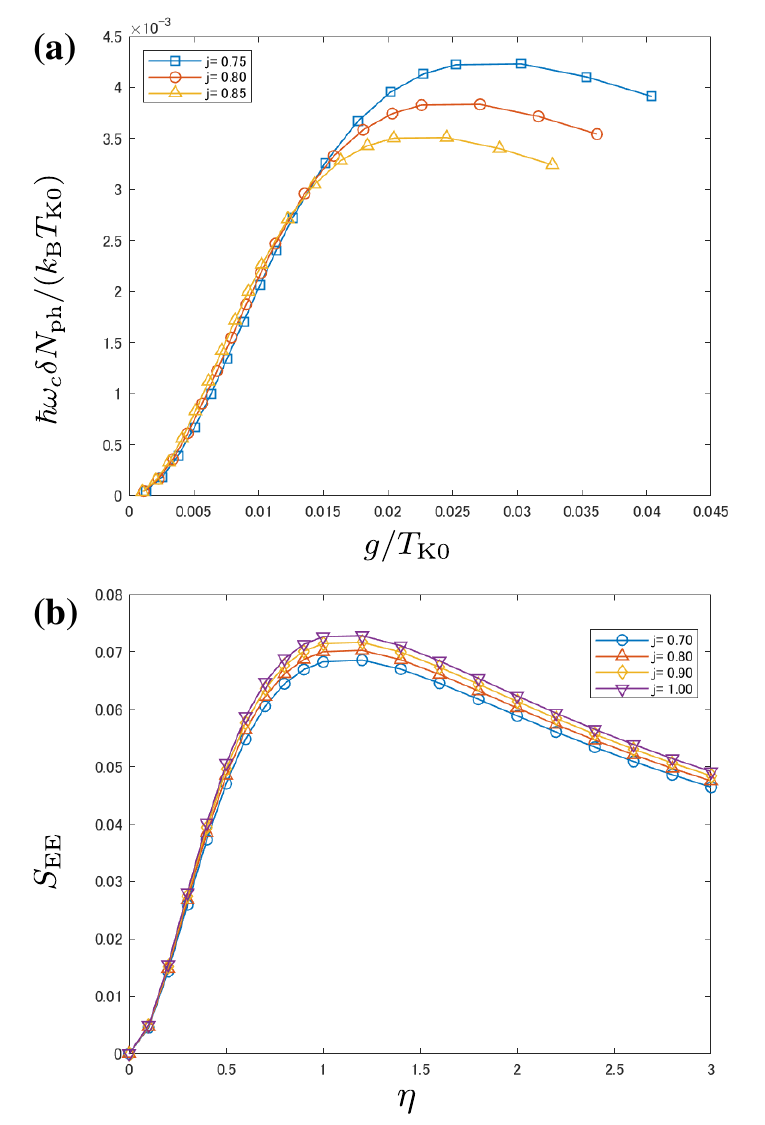}
	\caption{
		(a) The $g$ dependence of $\delta N_\mathrm{ph}(j, g)$, a component of the photon occupation number \eqref{eq:photon_number} that is induced by the Kondo effect. 
		The horizontal axis is normalized by the Kondo temperature $T_\mathrm{K0}$ at $\eta=0$, while the vertical axis plots the photon energy $\hbar\omega_c\delta N_\mathrm{ph}$ divided by the Kondo temperature $T_{{\rm K}0}$. 
		(b) Entanglement entropy between the electron system and cavity photons.
		The results are obtained for $N_\mathrm{e}=122$.		
		}
		\label{fig:photon}
\end{figure}

In this section, we discuss the influence of the Kondo effect on the quantum electromagnetic environment. Specifically, we consider the ground-state photon occupation number in the original Coulomb gauge, which is defined by (see Appendix~\ref{app:photon} for details)
\begin{align}
        N_\mathrm{ph}(j,g)&=\expval*{\hat{a}^\dagger \hat{a}}\notag\\
		&=\frac{\omega_c}{\Omega}\frac{\xi_g^2}{\hbar^2}\expval*{\hat{P}^2_\mathrm{e}}_\mathrm{GS}
        +\frac{1}{4}\left(\frac{\omega_c}{\Omega}+\frac{\Omega}{\omega_c}-2\right).
      \label{eq:photon_number}
\end{align}
The last term originates from the squeezing due to the $A^2$ term and exists even in the absence of the magnetic impurity. Meanwhile, the first term is the photon occupation induced by the Kondo effect and can be written as
\begin{align}
\delta N_\mathrm{ph}(j,g)&:=N_\mathrm{ph}(j,g)-N_\mathrm{ph}(0,g)\notag\\
&=\frac{\omega_c}{\Omega}\frac{\xi_g^2}{\hbar^2}\expval*{\hat{P}^2_\mathrm{e}}_\mathrm{GS}.\label{delph}
\end{align}
We note that the $\eta$ dependence of the Kondo-induced photon occupation $\delta N_\mathrm{ph}$ is mainly characterized by $\xi_g^2$, which scales as
\[\xi_g^2\propto\frac{\eta^2}{(1+2\eta^2)^{3/2}}.\] 
This quantity reaches a maximum value at $\eta=1$, and consequently, $\delta N_\mathrm{ph}(j,g)$ should exhibit a peak in the vicinity of $\eta= 1$ and monotonically decrease at larger $\eta$. 
Nevertheless, we note that the last term in Eq.~\eqref{eq:photon_number}, which is the squeezing contribution, always dominates over the Kondo-induced virtual photon contribution $\delta N_\mathrm{ph}$, and thus the total photon occupation $N_\mathrm{ph}$ still monotonically increases as a function of $\eta$.

The variational results of $\hbar\omega_c\delta N_\mathrm{ph}(j,g)$ are plotted in Fig.~\ref{fig:photon}(a). 
Interestingly, the photon energy $\hbar\omega_c\delta N_\mathrm{ph}$ exhibits the universality as a function of $g/T_{\mathrm{K}0}$, indicating the relation
\begin{align}
	\frac{\hbar\omega_c\delta N_\mathrm{ph}(j,g)}{k_\mathrm{B}T_\mathrm{K0}}=h\left(\frac{g}{T_\mathrm{K0}}\right)
\end{align}
with a scaling function $h(x)$ that satisfies $h(0)=0$.
The deviation from the universal curve in the deep strong coupling regime $\eta>1$ can be  understood from the fact that the photon excitation energy is renormalized to $\Omega$ and gets enhanced as $\Omega\propto \eta$ in this regime. Such a high photon excitation energy can lead to electron excitations far from the Fermi level where the dispersion relation can no longer be approximated as the linear one. 

The similar behavior can be also found in the entanglement entropy between the cavity photons and the electron system. 
One can obtain the entanglement entropy up to an order of $\order{\xi_g^2}$ as 
\begin{align}
	\hat{S}_\mathrm{EE}&=-\Tr_\mathrm{electron}\bigl[\hat{\rho}_0\log\hat{\rho}_0\bigr]=-\sum_{\sigma=\pm}\lambda_\sigma\log\lambda_\sigma,
	\label{eq:entanglement}\\
	\lambda_\pm&=\frac{1}{2}\left(1\pm\sqrt{1-4\xi_g^2\expval*{\hat{P}_\mathrm{e}^2}_\mathrm{GS}/\hbar^2}\right),
\end{align}
with $\hat{\rho}_0$ being the density matrix operator of the variational ground state in the Coulomb gauge (cf. Eq.~\eqref{gscoulomb}).
As shown in Fig.~\ref{fig:photon}(b), the $\eta$ dependence of this entanglement entropy is qualitatively similar to that of the Kondo-induced photon occupation~\eqref{eq:photon_number}.
Meanwhile, we note that the photon-electron entanglement shows the reduction in the deep strong coupling regime ($\eta>1$) and likely converges to zero in the limit $\eta\rightarrow\infty$.
This behavior arises from the fact that, as $\eta$ is increased, the $\hat{A}^2$ term in the original frame becomes  dominant, leading to the asymptotic decoupling of light and matter.

\section{Discussions}\label{sec:discuss}
\begin{figure}[tb]
	\centering
	\includegraphics[width=86mm]{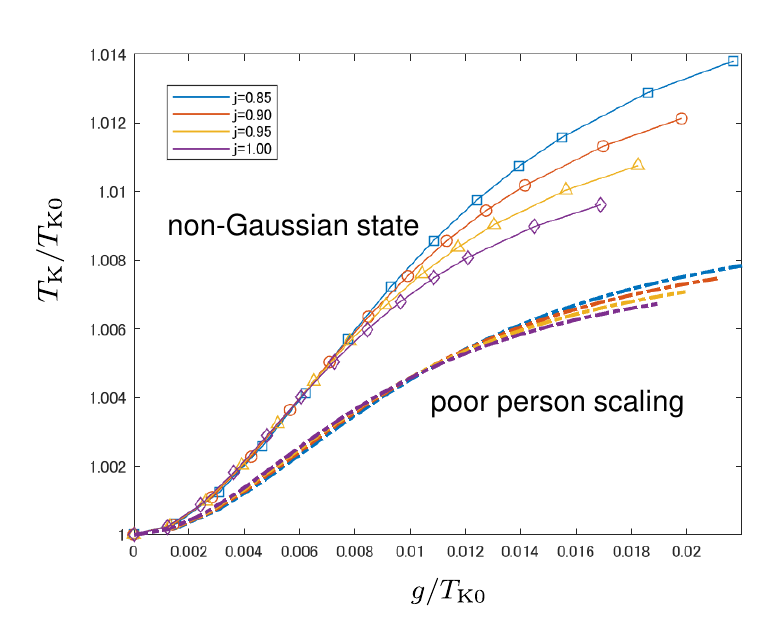}
	\caption{Comparison between the variational results and poor person's scaling. Dashed curves represent the enhancement of the Kondo temperature using the analytical expression~\eqref{eq:poorman_temperature} based on poor person's scaling. The solid curves represent the variational results obtained from the non-Gaussian states. The results are plotted  for $N_\mathrm{e}=70 $. Both variational and poor person's results are plotted as a function of $g/T_\mathrm{K0}$, indicating the universal relation at small $g/T_\mathrm{K0}$.}	\label{fig:tkondo_uni}
\end{figure}

We here provide a simple explanation of the variational results presented in the previous section on the basis of poor person's scaling~\cite{andersonPoorManDerivation1970}. 
After making several simplifications, the renormalization group equation for the scaling of the exchange interaction, including the cavity-mediated interaction, can be given by (see  Appendix~\ref{app:poor} for technical details)
\begin{align}
	\dd{j}=-\frac{\,\dd E_\Lambda}{\varepsilon-E_\Lambda(1-2\frac{g^2}{\Omega^2})}j^2,
\end{align}
where $E_\Lambda$ is the cutoff energy, and $\varepsilon$ is an excitation energy of electrons. 
This relation allows us to obtain the approximative analytical expression of the cavity-modified effective Kondo temperature as follows:
\begin{align}
	k_\mathrm{B} T_\mathrm{K}(\eta)&=k_\mathrm{B} T_{\mathrm{K}0}\exp[\frac{2}{N_\mathrm{e}}\frac{\eta^2}{1+2\eta^2}\frac{1}{j_0}].
	\label{eq:poorman_temperature}
\end{align}
In Fig.~\ref{fig:tkondo_uni}, we make a comparison between the variational results and the analytical result~\eqref{eq:poorman_temperature} obtained from poor person's scaling. Both results exhibit qualitatively similar behaviors and, in particular, indicate the universal relation of the cavity-modified Kondo temperature up to $\eta\sim 1$ (cf.~Eq.~\eqref{eq:temperature_uni}). 

In the setup we have studied in this paper, there exists only the single localized impurity embedded in the bulk metal, and the effect of the cavity confinement is expected to disappear when the number of conduction electrons $N_\mathrm{e}$ is taken to be infinite while keeping $\eta$ finite.   
In practice, however, metals contain a number of magnetic impurities with nonvanishing density. As far as the impurity concentration, $n_\text{imp} = N_\text{imp}/N_\mathrm{e}$, remains sufficiently small such that the RKKY interaction does not play an important role, we expect that our variational calculations should capture qualitative features of the cavity Kondo effect in such real materials including magnetic impurities. In principle, the Kondo cloud consists of a finite number of electrons in the vicinity of the localized impurity, and  the cavity-enhanced Kondo effect could be realized in each of individual magnetic impurities within the bulk.  It would be interesting to observe the cavity-induced shrinkage of such Kondo cloud dressed by virtual photons with the help of recent techniques \cite{v.borzenetsObservationKondoScreening2020,imObservationKondoCondensation2023} if at all possible.

Previous studies have indicated that the magnetic impurity system strongly interacting with bosonic degrees of freedom can give rise to the multi-channel Kondo effect. Examples include the systems with periodically driven classical light~\cite{ecksteinTwochannelKondoPhysics2017,quitoFloquetEngineeringMultichannel2023}or the electron-phonon interactions~\cite{diasdasilvaPhononassistedTunnelingTwochannel2009,mullerControlYuShibaRusinovStates2023}. 
In particular, in Ref.~\cite{lamacraftKondoPolaronsOneDimensional2008}, it has been discussed that the Hamiltonian of the Kondo polaron can contain a term proportional to $\hat{P}^2_\textrm{e}$, which gives rise to the multi-channel Kondo effect.
It merits further study to explore the possibility of realizing the multi-channel Kondo effect also in the setup with the cavity confinement, where the nonlocal $\hat{P}^2_\textrm{e}$ term can appear as shown in this paper; exploring such non-Fermi liquid feature induced by vacuum fluctuations of the cavity field could be one of the intriguing directions.
It would be also interesting to explore transport phenomena and real-time dynamics of the cavity Kondo effect, which can be done by using the real-time Gaussian variational calculations \cite{ashidaVariationalPrincipleQuantum2018,kanasz-nagyExploringAnisotropicKondo2018}, where
 the inclusion of photon excitations in the transformed frame could be important. From a broader perspective, our nonperturbative framework can be used to study a variety of quantum impurity problems influenced by structured quantum electromagnetic environment.

In conclusion, we have studied the influence of the cavity confinement on the Kondo effect in the ultrastrong coupling regime. Employing the two disentangling unitary transformations, we have obtained the effective model that describes the low-energy physics of the single magnetic impurity in a metal confined by the cavity. The leading contribution due to the cavity confinement can be captured by the nonlocal electron-electron interaction, leading to the mass renormalization akin to the polaronic mass enhancement. Consequently, the Kondo effect can be enhanced by the cavity confinement as confirmed in our variational calculations using fermionic Gaussian states. We also found that the ultrastrong coupling leads to universal scaling relations depending on the light-matter coupling strength $g$, which are unique to the cavity Kondo effect. We expect that our results should advance our understanding of cavity engineering of strongly correlated electronic phenomena. We hope that our work stimulates further studies in this direction. 

\begin{acknowledgments}
We are grateful to Atac Imamoglu, Kensuke Kobayashi, Kanta Masuki and Masaya Nakagawa, and Nato Tsuji for useful discussions. Y.A. acknowledges support from the Japan Society for the Promotion of Science (JSPS) through Grant No. JP19K23424 and from JST FOREST Program (Grant Number JPMJFR222U, Japan) and JST CREST (Grant Number JPMJCR23I2, Japan).  
\end{acknowledgments}

\appendix
\section{Derivation of the effective Hamiltonian}\label{app:exchange}
In this section,  we derive the low-energy effective Hamiltonian~\eqref{eq:ckondo} of the cavity Kondo effect, which includes the exchange interaction, by using perturbation theory.
We recall that the hybridization term is written as
\begin{align}
	&\hat{H}_\mathrm{hyb}=\frac{V_0}{\sqrt{N_\mathrm{e}}}\sum_{k,\sigma} \qty(e^{\xi_gk(\hat{b}^\dagger-\hat{b})}\hat{d}^{\dagger}_{\sigma}\hat{c}_{k\sigma}+\mathrm{h.c.}).
\end{align}
The terms in Eq.~\eqref{eq:anderson-cqed2} other than $\hat{H}_\mathrm{hyb}$ are the unperturbed Hamiltonian $\hat{H}_0$.
We consider the perturbative analysis by performing the expansion with respect to  the characteristic length,
\[\xi_g=g\sqrt{\frac{\hbar}{m}}\frac{1}{\Omega^{\frac{3}{2}}}
=\sqrt{\frac{\hbar}{m}}\frac{\eta}{\sqrt{N_\mathrm{e}\omega_c}(1+2\eta^2)^{\frac{3}{4}}},\]
which remains small over all the coupling strengths since it scales as $\xi_g^2\propto \order{N_{\mathrm{e}}^{-1}}$ at any $\eta$. 
Specifically, we expand the exponential function in $\hat{H}_\mathrm{hyb}$ with respect to $\xi_g k$ as follows:
\begin{align}
	\hat{H}_\mathrm{hyb}
	&=\hat{W}_0+\hat{W}_1+\hat{W}_2+\dots ,
\end{align}
where
\begin{align}
	\hat{W}_0&=\frac{V_0}{\sqrt{N_\mathrm{e}}}\sum_{k,\sigma}(\hat{d}^\dagger_\sigma \hat{c}_{k\sigma}+\hat{c}^\dagger_{k\sigma}\hat{d}_\sigma)\label{eq:V0}, \\
	\hat{W}_1&=\frac{\xi_gV_0}{\sqrt{N_\mathrm{e}}}\sum_{k,\sigma}k(\hat{b}^\dagger-\hat{b})(\hat{d}^\dagger_\sigma \hat{c}_{k\sigma}-\hat{c}^\dagger_{k\sigma}\hat{d}_\sigma)\label{eq:V1}, \\
	\hat{W}_2&=\frac{\xi_g^2 V_0}{2\sqrt{N_\mathrm{e}}}\sum_{k,\sigma}k^2(\hat{b}^\dagger-\hat{b})^2(\hat{d}^\dagger_\sigma \hat{c}_{k\sigma}+\hat{c}^\dagger_{k\sigma}\hat{d}_\sigma).\label{eq:V2}
\end{align}
To derive the effective  Hamiltonian, we consider the strong Coulomb interaction at the impurity site, i.e., $U\gg V_0$, and also assume that an impurity-site energy  is below the Fermi energy $\varepsilon_d<\varepsilon_f$.
In this parameter regime, doubly occupied state can be eliminated from the low-energy Hilbert space of interest, thereby leading to the single-magnetic moment state~\cite{andersonLocalizedMagneticStates1961,kondoResistanceMinimumDilute1964,schriefferRelationAndersonKondo1966}.

We begin by constructing the low-energy Hamiltonian by the second-order perturbation theory
\begin{align}
	\sum_{m}\frac{\bra{f} \hat{H}_\mathrm{hyb}\ket{m}\bra{m} \hat{H}_\mathrm{hyb}\ket{i}}{\expval*{\hat{H}_0}_i-\expval*{\hat{H}_0}_m}
	\label{eq:second_order}
\end{align}
with the initial state $\ket{i}$ and the final state $\ket{f}$. 
Firstly, only considering $\hat{W}_0$, one can get the usual Kondo exchange model,
\begin{align}
	H_{\mathrm{sd},0}&=\frac{J_0}{N_\mathrm{e}}\sum_{k,k'}\bigl[\hat{S}_\mathrm{imp}^{z}(\hat{c}^\dagger_{k'\uparrow}\hat{c}_{k\uparrow}-\hat{c}^\dagger_{k'\downarrow}\hat{c}_{k\downarrow})+\hat{S}_\mathrm{imp}^+\hat{c}^\dagger_{k\downarrow}\hat{c}_{k\uparrow}^\dagger\notag\\
	&+\hat{S}_\mathrm{imp}^-\hat{c}^\dagger_{k\uparrow}\hat{c}_{k\downarrow}^\dagger\bigr]+R_0\sum_{k,k',\sigma}\hat{c}^\dagger_{k\sigma}\hat{c}_{k'\sigma}
\end{align}
with the coupling constants
\begin{align}
	J_0&=V_0^2\qty[\frac{1}{\varepsilon_d+U}-\frac{1}{\varepsilon_d}],\\
	R_0&=\frac{V_0^2}{2N_\mathrm{e}}\qty[\frac{1}{\varepsilon_d+U}+\frac{1}{\varepsilon_d}].
\end{align}
We introduce the impurity spin operator $\hat{S}_\imp^\alpha$ for $\alpha=x,y,z$ and $\hat{S}_\imp^\pm=\hat{S}_\imp^x\pm i\hat{S}^y_\mathrm{imp}$.
Here, in the denominator, we assume $\varepsilon_f=0$ and neglect the kinetic energy term $\varepsilon_k=\hbar^2k^2/2m$ in the excitation energy near the Fermi surface.

Next, we consider the leading and next-leading terms $\hat{W}_1$ and $\hat{W}_2$ in 
Eq.~\eqref{eq:V1} and Eq.~\eqref{eq:V2}, respectively, which include virtual excitations and emissions of cavity photons. 
Since the hybridization process between the $d$ orbital and free electrons remains unchanged, the notable modifications from the conventional Kondo model appear in the perturbation's denominator, where the energy of the cavity field should be included, and in the term involving $\xi_g$ appearing as a product with $V_0$. 

Since $\hat{W}_1$ changes the number of photons, the change in the ground-state energy can originate from the second-order perturbation of $\hat{W}_1$. 
This modification accounts for the distinctive features introduced by the presence of virtual photon absorption and emission processes, differentiating the present model from the conventional Kondo model.
More specifically, the leading contribution from $\hat{W}_1$ is
\begin{align}
	\hat{H}_{\mathrm{sd},1}&=\frac{1}{N_\mathrm{e}}\sum_{k,k'}J_{1,kk'}\bigl[\hat{S}_\mathrm{imp}^{z}(\hat{c}^\dagger_{k'\uparrow}\hat{c}_{k\uparrow}-\hat{c}^\dagger_{k'\downarrow}\hat{c}_{k\downarrow})\notag\\
	&\quad+\hat{S}_\mathrm{imp}^+\hat{c}^\dagger_{k\downarrow}\hat{c}_{k\uparrow}^\dagger+\hat{S}_\mathrm{imp}^-\hat{c}^\dagger_{k\uparrow}\hat{c}_{k\downarrow}^\dagger\bigr]\notag\\ &\quad
	+\sum_{k,k',\sigma}R_{1,kk'}\hat{c}^\dagger_{k\sigma}\hat{c}_{k'\sigma},\\
	J_{1,kk'}&=V_0^2\xi_g^2\qty[\frac{1}{\varepsilon_d+U+\hbar \Omega}-\frac{1}{\varepsilon_d-\hbar\Omega}]kk',   \\
	R_{1,kk'}&=\frac{V_0^2}{2N_\mathrm{e}}\xi_g^2\qty[\frac{1}{\varepsilon_d+U+\hbar\Omega}+\frac{1}{\varepsilon_d-\hbar\Omega}]kk'.
\end{align}
The exchange interaction appears in a form that depends on the wavevector $k$.

Meanwhile, the leading contribution from $\hat{W}_2$ comes from the first-order perturbation of   $\hat{W}_2$, which reads as
\begin{align}
	\hat{H}_{\mathrm{sd},2}&=\sum_{k,k'}J_{2,kk'}\bigl[\hat{S}_\mathrm{imp}^{z}(\hat{c}^\dagger_{k'\uparrow}\hat{c}_{k\uparrow}-\hat{c}^\dagger_{k'\downarrow}\hat{c}_{k\downarrow})\notag\\ &\quad+\hat{S}_\mathrm{imp}^+\hat{c}^\dagger_{k\downarrow}\hat{c}_{k\uparrow}^\dagger
	+\hat{S}_\mathrm{imp}^-\hat{c}^\dagger_{k\uparrow}\hat{c}_{k\downarrow}^\dagger\bigr]\notag\\ &\quad
	+\sum_{k,k',\sigma}R_{2,kk'}\hat{c}^\dagger_{k\sigma}\hat{c}_{k'\sigma},\\
	J_{2,kk'}&=-\frac{1}{2}V_0^2\xi_g^2\qty[\frac{1}{\varepsilon_d+U}-\frac{1}{\varepsilon_d}](k^2+k^{'2}),   \\
	R_{2,kk'}&=-\frac{V^2_0}{4}\xi_g^2\qty[\frac{1}{\varepsilon_d+U}+\frac{1}{\varepsilon_d}](k^2+k^{'2}).
\end{align}
When we consider the symmetric case $\varepsilon_d=-U/2$,  the interactions become $J_0=4V^2_0/U,R_0=0$ and
\begin{align}
	J_{1,kk'}
	&=\frac{\xi_g^2 kk'}{1+2J_0\hbar \Omega/V^2_0}J_0\sim\xi_g^2kk' J_0,
\end{align}
where we neglect the second term in the denominator since $J_0\hbar\Omega/V_0^2=4\hbar\Omega/U\ll 1$ as inferred from $U\gg\hbar\Omega=\hbar\omega_c\sqrt{1+2\eta^2}$.

Finally, we arrive at the following Hamiltonian
\begin{align}
        \hat{H}_\mathrm{cK}&=\frac{\hbar^2}{2m}\sum_{k,\sigma}k^2\hat{c}^\dagger_{k\sigma}\hat{c}_{k\sigma}-\frac{\hbar^2g^2}{m\Omega^2}\hat{P}_\mathrm{e}^2\notag\\ &\quad
		+\frac{1}{2N_\mathrm{e}}\sum_{k,k',\sigma,\sigma'}J_{kk'}\,\hat{c}^\dagger_{k\sigma}(\vec{\sigma})^{\sigma\sigma'}\hat{c}_{k'\sigma'}\cdot\vec{S}_\mathrm{imp}, \\
        J_{kk'}&=\qty(1-\frac{1}{2}\xi_g^2 (k-k')^2)J_0.\label{jkk}
    \end{align}
We note that the spin-unrelated scattering $R$ vanishes.
While we assume the symmetric model $\varepsilon_d=-U/2$ here, we expect that our conclusions remain qualitatively similar in other cases since only the magnitude of $J_{kk'}$ can be slightly affected, and a nonzero spin-independent scattering $R$ only modifies the chemical potential.

The second term $\frac{1}{2}\xi_g^2(k-k')^2$ of $J_{kk'}$ in Eq.~\eqref{jkk} gives a contribution which is at most $\sim 2\xi_g^2 k_f$. We recall that $\xi_g^2$ scales as $\order{N_\mathrm{e}^{-1}}$.
In the ultrastrong coupling regime with the collective enhancement, $\eta=\sqrt{N_\mathrm{e}}g/\omega_c$ can be $\eta\sim\order{1}$.
Even when $\xi_g$ takes its maximum value around $\eta=1$, the contribution $2\xi_g^2 k_f^2$ remains small such that the $k$ and $k'$ dependencies in Eq.~\eqref{jkk} can be neglected. Indeed, we have numerically checked that the inclusion of  the momentum-dependent term in $J_{kk'}$ only leads to minuscule changes in the results and does not affect the conclusions of this paper.
We thus obtain the effective Hamiltonian~\eqref{eq:ckondo} in the main text. 

\section{Derivation of the photon occupation and entanglement entropy}\label{app:photon}
In this section, we provide the expressions of the photon occupation number in Eq.~\eqref{eq:photon_number} and the entanglement entropy in Eq.~\eqref{eq:entanglement}.
First of all, using the relation in the Bogoliubov transformation~\eqref{eq:bogoliubov}, the expectation value of the photon number of the $\hat{a}$ operator in the Coulomb gauge can be rewritten as 
\begin{align}
	N_\mathrm{ph}(j,g)&=\expval*{\hat{a}^\dagger \hat{a}}\notag\\
	&=\frac{\omega_c}{\Omega}(1+\eta^2)\expval*{\hat{b}^\dagger\hat{b}}-\frac{\eta^2}{2}\frac{\omega_c}{\Omega}\expval*{\hat{b}^2+\hat{b}^{\dagger 2}}\notag\\
	&\quad +\frac{1}{4}\left(\frac{\omega_c}{\Omega}+\frac{\Omega}{\omega_c}-2\right).
\end{align}
Using the expressions of the variational ground state 
$\ket{\psi}=\hat{U}_\mathrm{AD}\hat{U}_\mathrm{K}\qty(\ket{0}_\mathrm{ph}\ket{\sigma_x}_\imp \ket{\psi_\mathrm{GS}})$ in the Coulomb gauge, where $\ket{\psi_\mathrm{GS}}$ is a fermionic Gaussian state, we obtain the expectation values
\begin{align}
	\expval*{\hat{b}^\dagger\hat{b}}&=\frac{\xi_g^2}{\hbar^2}\expval*{\hat{P}^2_\mathrm{e}}_\mathrm{GS}, \\
	\expval*{\hat{b}^2+\hat{b}^{\dagger 2}}&=\frac{2\xi_g^2}{\hbar^2}\expval*{\hat{P}^2_\mathrm{e}}_\mathrm{GS}.
\end{align}
We here note that the vacuum state is defined in terms of the $\hat{b}$ operator, i.e., $\hat{b}\ket{0}_\mathrm{ph}=0$, and the unitary transformation acts as $\hat{U}_\mathrm{AD}^\dagger \hat{b}\hat{U}_\mathrm{AD} = \hat{b}+\xi_g \hat{P}_\mathrm{e}/\hbar$.
As a result, we get
\begin{align}
	N_\mathrm{ph}(j,g)=\frac{\xi_g^2}{\hbar^2}\frac{\omega_c}{\Omega}\expval*{\hat{P}^2_\mathrm{e}}_\mathrm{GS}+\frac{1}{4}\left(\frac{\omega_c}{\Omega}+\frac{\Omega}{\omega_c}-2\right).
\end{align} 

To calculate the entanglement entropy between the cavity electromagnetic field and the total electron system,  including the impurity and conduction electrons, we define a reduced density operator by tracing out the electron part 
\begin{align}
	\hat{\rho}_\mathrm{ph}=\Tr_\mathrm{e}\bigl[\ket{\psi}\bra{\psi}\bigr].
\end{align}
We use the Fock space basis $\{\ket{n}_\mathrm{ph}\}$ with respect to the $\hat{b}$ photon field, the partial trace results in 
\begin{align}
	\hat{\rho}_\mathrm{ph}=\sum_{n,m=0} \,\frac{\ket{n}_\mathrm{ph}\bra{m}}{\hbar^{n+m}\sqrt{n!m!}}\xi_{g}^{n+m}\expval*{\hat{P}_\mathrm{e}^{n+m}e^{-\xi_g^2\hat{P}^2_\mathrm{e}}}_\mathrm{GS}.
\end{align}
The leading contribution in terms of $\xi_g \hat{P}_{\rm e}$ to the  entanglement entropy as a von Neumann entropy gives
\begin{align}\label{eq:ee_app}
	S_\mathrm{EE}&=-\Tr[\hat{\rho}_\mathrm{ph}\log\hat{\rho}_\mathrm{ph}] \\
	&\simeq -\sum_{\sigma=\pm}\lambda_\sigma\log\lambda_\sigma,
\end{align}
where
\begin{align}
	\lambda_\pm&=\frac{1}{2}\left(1\pm \sqrt{1-4\xi^2_g\expval*{\hat{P}^2_\mathrm{e}}_\mathrm{GS}/\hbar^2}\right).
\end{align}
While the entanglement entropy in Eq.~\eqref{eq:ee_app} is not sensitive to a choice of the photon basis, it does so in practice when truncating the photon number to evaluate its approximate value. Our results indicate that the convergence of this approximation is much better when considering the Fock space of the $\hat{b}$ photon than that of the   $\hat{a}$ photon.

\section{Poor person's scaling}\label{app:poor}
Here we provide the details of poor person's scaling, which has been originally developed by Anderson~\cite{andersonPoorManDerivation1970}. 
Consider a cut-off energy $\pm E_\Lambda$ that is away from the Fermi energy at $\varepsilon_f=0$.
When this cut-off energy is slightly modified by the amount $\Delta E>0$ such that $E'_\Lambda=E_\Lambda-\Delta E$, the change of the potential energy is
\begin{align}
	\dd{\hat{V}}=\hat{V} \hat{\mathrm{P}}_{\Delta E}\hat{G}_0 \hat{V}+\order{\Delta E^2},
\end{align}
where $\hat{\mathrm{P}}_{\Delta E}$ is the projection operator onto a space $E_\Lambda>|\varepsilon|>E_\Lambda-\Delta E$ and $\hat{G}_0(\varepsilon)=(\varepsilon-\hat{H}_0)^{-1}$ is the unperturbed Green's function. 
The potential displacement $\dd \hat{V}$ is the operator on the states that belong to the subspace of $1-\hat{\mathrm{P}}_{\Delta E}$.

We define the unperturbed Hamiltonian $\hat{H}_0$ and the potential energy $\hat{V}$ as follows:
\begin{align}
	\hat{H}_0&=\sum_k \varepsilon_{k}\hat{c}^\dagger_{k\sigma}\hat{c}_{k\sigma}-\frac{\hbar^2g^2}{m\Omega^2}\biggl(\sum_{k,\sigma}k\hat{c}^\dagger_{k\sigma}\hat{c}_{k\sigma}\biggr)^2,\\
	\hat{V}&=J\vec{S}_\mathrm{imp}\cdot\vec{s}(0)\notag\\
	&=\frac{J}{2N_\mathrm{e}}\vec{S}_\mathrm{imp}\cdot\sum_{k,k',\sigma,\sigma'}\hat{c}_{k\sigma}^\dagger\vec{\sigma}^{\sigma\sigma'}\hat{c}_{k'\sigma'}.
\end{align}

\begin{widetext}
The change in the potential can be calculated as
\begin{align}
	\dd{\hat{V}}&=\frac{1}{4N_\mathrm{e}}J^2\sum_{k,k',\sigma,\sigma'}^{|\varepsilon_{k},\varepsilon_{k'}|<E_\Lambda-\Delta E}\sum_{\substack{p,\xi \\ E_\Lambda-\Delta E<|\varepsilon_p|<E_\Lambda}}
	\Bigl[\vec{S}_\mathrm{imp}\cdot(\vec{\sigma}^{\sigma\xi})\Bigr]\Bigl[\vec{S}_\mathrm{imp}\cdot(\vec{\sigma}^{\xi\sigma'})\Bigr]\notag\\
	&\times\Biggl[\frac{1}{\varepsilon-(\varepsilon_f+\varepsilon_p-\varepsilon_{k'}-v_{p,k'})}\hat{c}^\dagger_{k\sigma}\hat{c}_{p\xi}\hat{c}^\dagger_{p\xi}\hat{c}_{k'\sigma'} 
	 +\frac{1}{\varepsilon-(\varepsilon_f+\varepsilon_{k'}-\varepsilon_p-v_{p,k'})}\hat{c}^\dagger_{p\xi}\hat{c}_{k\sigma}\hat{c}^\dagger_{k'\sigma'}\hat{c}_{p\xi}\Biggr],
\end{align}
where we introduce the variable $v_{p,k'}=\frac{\hbar^2g^2}{m\Omega^2}(p^2+k^{'2}-2pk')$ for the sake of notational simplicity. 
We note that the terms proportional to $\sum_{q,\tau}q\hat{c}^\dagger_{q\tau}\hat{c}_{q\tau}$ and $\sum_{q,q',\tau,\tau'}qq'\hat{c}^\dagger_{q\tau}\hat{c}_{q\tau}\hat{c}^\dagger_{q'\tau'}\hat{c}_{q'\tau'}$ vanish in a Fermi state. 
Considering sufficiently low temperatures, the wavenumbers $k,k'$ can be treated as the variables denoting the excitations near the Fermi surface. 
Also, high-energy momentum $p$ can be taken as an excitation that is nearly at the energy cutoff, which is sufficiently far from the Fermi surface, and the energy slice $\Delta E$ is sufficiently small.
Therefore, we assume that the $p$ dependence of the $v_{p,k'}$ can be neglected, as $v_{k'}=v_{p,k'}\bigl|_{|p|=k_\Lambda}$.

Integrating out the virtual states, we get
\begin{multline}
	\dd{\hat{V}}=\frac{1}{4N_\mathrm{e}}J^2\sum_{k,k',\sigma,\sigma'}^{|\varepsilon_{k},\varepsilon_{k'}|<E_\Lambda-\Delta E}\sum_{\xi}\Bigl[\vec{S}_\mathrm{imp}\cdot(\vec{\sigma}^{\sigma\xi})\Bigr]\Bigl[\vec{S}_\mathrm{imp}\cdot(\vec{\sigma}^{\xi\sigma'})\Bigr]\\
	 \times \left[ \frac{\rho_F \Delta E}{\varepsilon-E_\Lambda+\varepsilon_{k'}+v_{k'}}\hat{c}^\dagger_{k\sigma}\hat{c}_{k'\sigma'}
	+\frac{\rho_F\Delta E}{\varepsilon-\varepsilon_{k'}-E_\Lambda+v_{k'}}\hat{c}_{k\sigma}\hat{c}^\dagger_{k'\sigma'}\right].
	\label{eq:dv_totyuu}
\end{multline}
Using the relation $\sigma^{\alpha}\sigma^{\beta}=\delta_{\alpha\beta}+i\varepsilon_{\alpha\beta\gamma}\sigma^{\gamma}$ with the Levi-Civita symbol $\varepsilon_{\alpha\beta\gamma}$, we can take the exchange interactions out of the expression \eqref{eq:dv_totyuu} such that 
\begin{align}
	\dd{\hat{V}}
	&=-\frac{1}{4}J^2\, \sum_{k,k'}\left[ \frac{\rho_F\Delta E}{\varepsilon-E_\Lambda+\varepsilon_{k'}+v_{k'}}
	+\frac{\rho_F\Delta E}{\varepsilon-\varepsilon_{k'}-E_\Lambda+v_{k'}}\right]
	\vec{S}_\mathrm{imp}\cdot \hat{c}_{k\sigma}(\vec{\sigma})^{\sigma\sigma'}\hat{c}^\dagger_{k'\sigma'} .
\end{align}
\end{widetext}
We assume that the electron excitations are only being in the vicinity of the Fermi energy $\varepsilon_F=0$. The denominators can then be simplified as $\varepsilon_k\simeq\varepsilon_{k'}\simeq0$ and $v_{k'}\simeq 2E_\Lambda g^2/\Omega^2$, leading to 
\begin{align}
	\dd{\hat{V}}
	&=-\frac{J^2\rho_F\Delta E}{\varepsilon-E_\Lambda(1-2\frac{g^2}{\Omega^2})}\,\vec{S}_\mathrm{imp}\cdot \vec{S}(0).
\end{align}
Thus, the scaling of the exchange interaction becomes
\begin{align}
	\dd{J}=-\frac{\rho_F\,\dd E_\Lambda}{\varepsilon-E_\Lambda(1-2\frac{g^2}{\Omega^2})}J^2.
\end{align}
Taking the limit $\varepsilon\rightarrow0$ with the initial conditions $J(E_{\Lambda_0})=J_0,\,E_{\Lambda_0}=D$, we can integrate this equation to get
\begin{align}
	\frac{1}{\rho_FJ(E_\Lambda)}=\frac{1}{\rho_F J_0}+\Bigl(1-2\frac{g^2}{\Omega^2}\Bigr)^{-1}\log\frac{E_\Lambda}{D}.
\end{align}
The cut-off energy $E_\Lambda$ at which $J(E_\Lambda)$ goes to infinity can be defined as the Kondo temperature $T_\mathrm{K}$ as follows:
\begin{align}
	k_\mathrm{B} T_\mathrm{K}(\eta)&=k_\mathrm{B} T_{\mathrm{K}0}\exp[\frac{2g^2}{\Omega^2}\frac{1}{\rho_F J_0}] \\
	&=k_\mathrm{B} T_{\mathrm{K}0}\exp[\frac{2}{N_\mathrm{e}}\frac{\eta^2}{1+2\eta^2}\frac{1}{\rho_F J_0}],
\end{align}
where $k_\mathrm{B}T_{\mathrm{K}0}=De^{-1/(\rho_F J_0)}$ is the Kondo temperature at $\eta=0$.
This formula provides the cavity-enhanced Kondo temperature in Eq.~\eqref{eq:poorman_temperature} in the main text.

\bibliography{cavityKondo_manuscript}

\end{document}